\documentclass[aps,prl,reprint,superscriptaddress,floatfix,nofootinbib]{revtex4-2}
\usepackage[usenames, dvipsnames]{color}
\usepackage{amsmath}
\usepackage{amsfonts}
\usepackage{comment}
\usepackage{hyperref}
\usepackage{ragged2e}
\usepackage[caption=false]{subfig}
\hypersetup{colorlinks=true, linkcolor=Blue, citecolor=Blue, filecolor=Blue, urlcolor=Blue}
\usepackage{graphicx}
\usepackage{physics}
\usepackage[normalem]{ulem}

\begin{document}
\title{Robust continuous symmetry breaking and multiversality in the chiral Dicke model}
\author{Nikolay Yegovtsev}
\affiliation{Department of Physics and Astronomy and IQ Initiative, University of Pittsburgh, Pittsburgh, Pennsylvania 15260, USA}
\author{Sayan Choudhury}
\email{sayanchoudhury@hri.res.in}
\affiliation{Harish-Chandra Research Institute, Chhatnag Road, Jhunsi,
Prayagraj 211 019, India}
\affiliation{Homi Bhabha National Institute, Training School Complex,
Anushakti Nagar, Mumbai 400 094, India}
\author{W. Vincent Liu}
\email{wvliu@pitt.edu}
\affiliation{Department of Physics and Astronomy and IQ Initiative, University of Pittsburgh, Pittsburgh, Pennsylvania 15260, USA}

\date{\today}
	
\begin{abstract}
The Dicke model (DM) serves as a paradigm for understanding collective light-matter interactions. We introduce the chiral Dicke model, a generalization where an atomic ensemble couples to a two-mode cavity via chiral interactions. Unlike the standard DM, the chiral DM is endowed with an inherent continuous $U(1)$ symmetry associated with angular momentum conservation. The ground-state phase diagram and the associated quantum phase transitions are charted out, revealing a $U(1)$-broken superradiant phase that spans a broad parameter space. We demonstrate that the spectrum of quantum fluctuations is highly tunable in both the symmetric and broken phases. Strikingly, our calculations reveal that the system exhibits `multiversality', where distinct universality classes govern the transition between the same two phases. In particular, along a special line in parameter space, the dynamical critical exponent for the normal-superradiant phase transition changes from $z\nu=1$ to $z\nu=1/2$. Our work establishes the chiral Dicke model as a powerful platform to realize novel quantum phases and multiversal critical phenomena in light-matter coupled systems. 
\end{abstract}

\maketitle

The quantum phases and dynamical properties of collective quantum systems have attracted considerable attention in recent years~\cite{defenu2024out,defenuRMP,sinha2024classical}. These systems serve as a natural platform, both for the generation of metrologically useful entangled states~\cite{kitagawa1993squeezed,ma2009fisher,jin2009spin,ma2011quantum,liu2011spin,munoz2023phase,carrasco2024dicke,gangopadhay2025counterdiabatic,reilly2024speeding,biswas2025discrete,biswas2025floquet}, and for the realization of novel non-equilibrium phases of matter~\cite{russomanno2017floquet,pizzi2021higher,lyu2020eternal,lerose2025theory}. In this context, an extensive number of investigations have focused on the Dicke model (DM) that describes an ensemble of $N$ two-level atoms coupled to a single quantized mode of a cavity~\cite{DickeOG,garraway2011dicke,kirton2019introduction,roses2020dicke,frisk2019ultrastrong}. This model exhibits a quantum phase transition (QPT), from the normal state to the superradiant $\mathbb{Z}_2$ symmetry-breaking state, when the strength of the light-matter interaction exceeds a critical value~\cite{HeppLieb, WangHioe}. Intriguingly, this ground-state QPT is accompanied by a spectral transition from integrability to chaos~\cite{emary&brandesprl,emary&brandespre,lewis2019unifying,das2022revisiting}. Consequently, the DM has provided a fertile platform for exploring dynamical signatures of quantum chaos~\cite{song2009spin,bastarrachea2015chaos,lobez2016entropy,pawar2025comparative,wang2020statistical}. Notably, the DM has been realized in a variety of platforms including cavity QED~\cite{baumann2010dicke,klinder2015dynamical,baden2014realization}, circuit QED~\cite{mlynek2014observation,tomonaga2025spectral}, trapped ion processors~\cite{safavi2018verification,gilmore2021quantum,bullock2026quantum}, and magnon-spin coupled systems~\cite{kim2025observation,baydin2025perspective}.

The experimental realization of the DM has led to theoretical investigations into generalizations of the DM, that can host a richer array of both equilibrium and non-equilibrium quantum phases \cite{dimer2007proposed,nonreciprocalD,keeling2010collective,bhaseen2012, unbalanceddicke2020,Mivehvar2024,alavirad2019scrambling,buijsman2017nonergodicity,das2023phase,zhu2024quantum}. Interestingly, these studies have pointed to the possibility of enhancing the symmetry of the model from the discrete $\mathbb{Z}_2$ symmetry to a continuous $U(1)$ symmetry by tuning the atom-photon coupling strengths ~\cite{baksic2014controlling}. This leads to the possibility of a chiral $U(1)$-broken superradiant phase However, this $U(1)$ symmetry generally emerges at a fine-tuned set of parameter values, and consequently, the chiral $U(1)$-broken superradiant state is very fragile~\cite{soriente2018dissipation}. Two questions naturally arise in this context: (a) Is it possible to realize a robust continuous-symmetry-broken phase in a generalized DM? and (b) Is it possible to promote the standard non-chiral DM to a chiral DM? 

In this Letter, we answer these questions affirmatively by introducing the $U(1)$-symmetric {\emph {chiral}} DM. This model constitutes the many-atom generalization of the two-mode chiral Rabi model that describes the coupling of the circularly polarized transition dipole of a two-level system (TLS) to the circularly polarized modes of an optical cavity in the ultrastrong coupling regime~\cite{chiral2019}. We demonstrate that this model hosts two distinct robust phases: the $U(1)$-symmetric normal phase and the $U(1)$-broken superradiant phase. We further analyze the spectrum of the quantum fluctuations in these phases and establish their parametric tunability. Notably, the dynamical critical exponent $z \nu$ for the normal-superradiant quantum phase transition can take two values, $1$ and $1/2$ in different regions of the critical line, thereby establishing the `multiversal' nature of the transition. Our results indicate that the chiral DM can serve as a fertile platform for tunably realizing novel phases and phase transitions. 

\begin{figure}[t]
        \includegraphics[width = 0.47\textwidth]{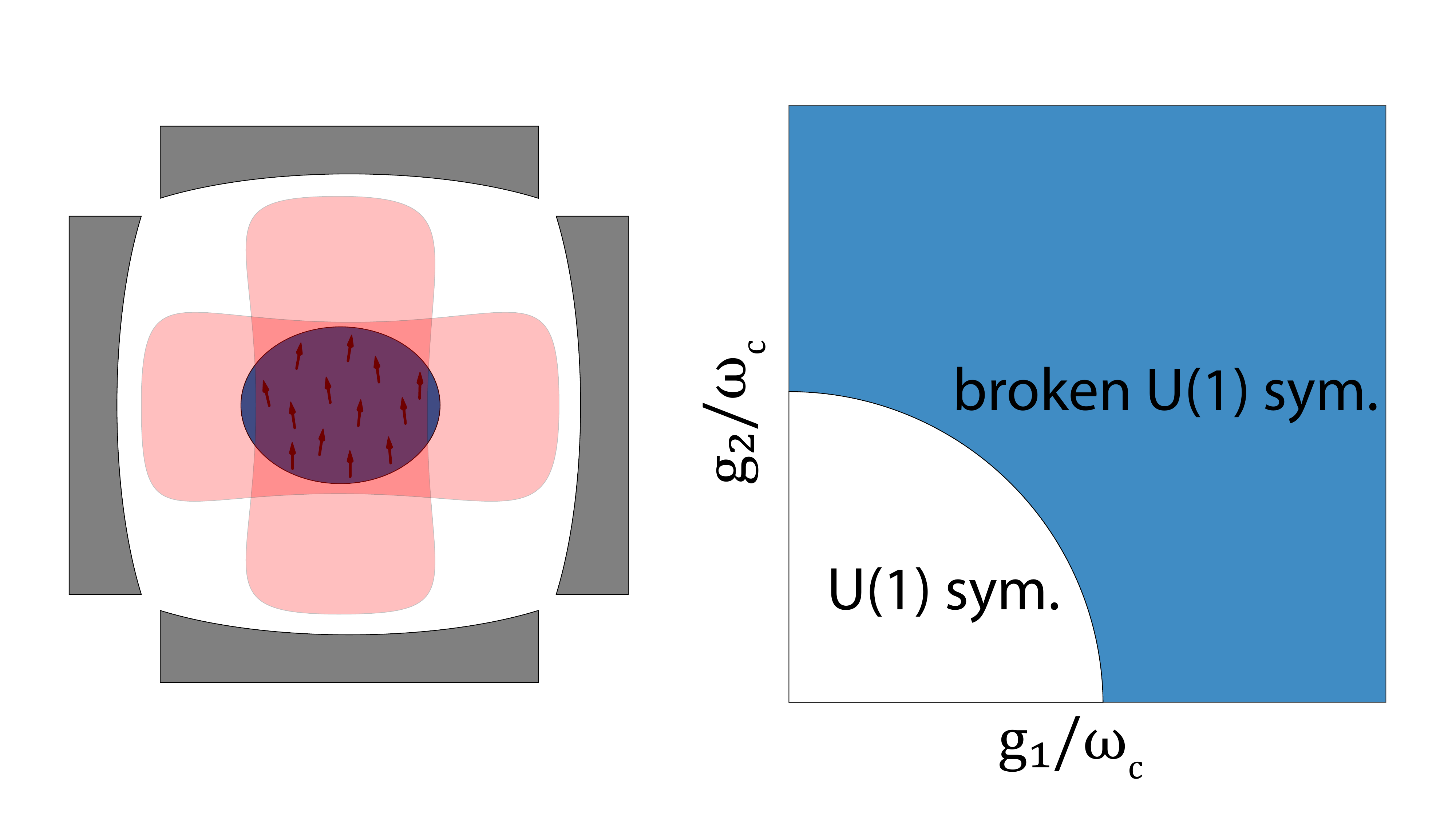}
    \caption{(Left: A schematic illustration of chiral Dicke model described in Eq.~\eqref{eq:Hcdm}. Right: The ground-state phase diagram of this system showing the existence of the $U(1)$-symmetric normal phase (white region), and the $U(1)$-broken superradiant phase. The location of the phase boundary is given by Eq.~\eqref{eq:phaseboundary}. }
    \label{fig:1}
\end{figure}

{\emph {Model:}} We now proceed to introduce the chiral DM which describes the dynamics of $N$ TLS atoms coupled via chiral light-matter interactions with two degenerate cavity modes of frequency $\omega_c>0$. Crucially, these atoms separately couple to the two modes through co-rotating and counter-rotating terms with strength $g_1$ and $g_2$, respectively:
\begin{equation}
\label{eq:Hcdm}
\begin{split}
H_{\rm CDM} = &\, \omega_c(a_1^\dagger a_1+a_2^\dagger a_2)+\omega_zS^{z}+ US^{z}(a_1^\dagger a_1+a_2^\dagger a_2)\\
& +\frac{g_1}{\sqrt{N}}(a_1S^{+}+a_1^\dagger S^{-}) +\frac{g_2}{\sqrt{N}}(a_2S^{-}+a_2^\dagger S^{+}),   
\end{split}
\end{equation}
where $S^{\pm} = \frac{1}{2} \sum_{j=1}^N (\sigma_j^{\rm x} \pm i \sigma_j^{\rm y})$ and $\{\omega_z,g_1,g_2\} >0 $. The route to experimentally realize $H_{\rm CDM}$ for a single TLS ($N=1$) has been discussed in Ref.~\cite{chiral2019}. A schematic illustration of this system is given in Fig.~\ref{fig:1}. 

The defining characteristic of $H_{\rm CDM}$ that distinguishes it from other generalized DMs is the chiral $U(1)$ symmetry: $a_1\to e^{i\theta}a_1$, $a_2\to e^{-i\theta}a_2$, $S^{\pm} \to e^{\mp i\theta}S^{\pm}$. This symmetry reflects the conservation of the total angular momentum $L^{z} = a_1^\dagger a_1-a_2^\dagger a_2+S^{z}$. 
We stress that this $U(1)$-symmetry and $L^{z}$ conservation persists for arbitrary values of couplings $g_1$ and $g_2$ in the chiral DM. This is in stark contrast to other two-mode Dicke models, where the $U(1)$ symmetry emerges only at fine-tuned parameter regimes~\cite{hiddenU1}. We now proceed to investigate the nature of the phases and phase transitions in this system.

\emph{Phase diagram:} We begin our analysis by obtaining the ground state phase diagram by employing mean-field theory~\cite{emary&brandesprl, emary&brandespre}. To this end, we first apply the Holstein-Primakoff transformation: $S^{+} = a_3^\dagger\sqrt{N-a_3^\dagger a_3}$, $S^{-} = \left(\sqrt{N-a_3^\dagger a_3}\right)a_3$, $S^{z} = (a_3^\dagger a_3-N/2)$, where the bosonic operator $a_3$ obeys the standard commutation relation, $[a_3,a_3^\dagger] = 1$.  We then shift all operators by their mean-field values, $\tilde{a}_i = \langle a_i \rangle$, such that $a_1 \to \alpha_1 +\tilde{a}_1$, $a_2 \to \alpha_2 +\tilde{a}_2$, $a_3 \to \alpha_3 + \tilde{a}_3$. The normal and superradiant phase is characterized by $\alpha_i=0$ and $\alpha_i \propto \sqrt{N}$ respectively.

The ground-state energy in the mean-field approximation thus takes the form:
\begin{equation}
\begin{split}
&E_G =\tilde{\omega}_c(|\alpha_1|^2+|\alpha_2|^2) +\omega_z|\alpha_3|^2 +U|\alpha_3|^2(|\alpha_1|^2+|\alpha_2|^2)\\
&+\frac{\sqrt{N-|\alpha_3|^2}}{\sqrt{N}}\left[  g_1(\alpha_1\alpha_3^*+\alpha_1^*\alpha_3)+g_2(\alpha_2\alpha_3+\alpha_2^*\alpha_3^*)\right],   
\end{split}
\end{equation}
where $\tilde{\omega}_c = \omega_c-UN/2$. We note that in this expression, we have only retained terms proportional to $N$, and neglected constant terms. The chiral U(1) symmetry is realized as $\alpha_1\to e^{i\theta}\alpha_1$, $\alpha_2\to e^{-i\theta}\alpha_2$, and $\alpha_3 \to e^{i\theta}\alpha_3$. The standard minimization procedure with respect to $\{\alpha_1^{*}, \alpha_2^{*}\}$ yields an expression for the ground state energy that is invariant under the phase rotations of $\alpha_3$: 
\begin{equation}
\label{eq:Heff}
E_G (\alpha_3, \alpha_3^{*}) = \omega_z|\alpha_3|^2-(g_1^2+g_2^2)\frac{|\alpha_3|^2(N-|\alpha_3|^2)}{N(\tilde{\omega}_c+U|\alpha_3|^2) }. 
\end{equation}
We perform our computations in the regime $4\omega_c^2>U^2N^2$ to ensure that the ground state energy is a bounded continuous function for the physically admissible values of $|\alpha_3|<N$.

We now perform a further minimization procedure with respect to $\alpha_3^{*}$ that yields two solutions, one  corresponding to the normal state ($\alpha_3=0$) and another corresponding to the superradiant state:
\begin{equation}
\label{eq:op}
\alpha_3 = \sqrt{\frac{\tilde{\omega}_c}{U}\left(\sqrt{\frac{\tilde{\omega}_c+UN}{\tilde{\omega}_c+\tilde{\mu}UN}}-1\right)}e^{i\theta},    
\end{equation}
where $\tilde{\mu} = \omega_z\tilde{\omega}_c/(g_1^2+g_2^2)$. We observe that the amplitude of the superradiant solution is only real above the critical coupling $\tilde{\mu}<1$. A more detailed analysis reveals that the trivial state is a local minimum of the energy below the critical coupling $\sqrt{g_1^2+g_2^2}<\sqrt{\omega_z\tilde{\omega}_c}$, and becomes a local maximum, while the superradiant state is an energy minimum when $\sqrt{g_1^2+g_2^2}>\sqrt{\omega_z\tilde{\omega}_c}$. Thus, the transition between these two phases occur when
\begin{equation}
\label{eq:phaseboundary}
\sqrt{g_1^2+g_2^2} = g_{c} = \sqrt{\omega_z\tilde{\omega}_c},    
\end{equation}
We observe that $\alpha_3$ in Eq.~\eqref{eq:op} continuously transitions from $0$ to a finite value at $g=g_{c}$, and can serve as the order parameter. The mean-field phase diagram is presented in Fig.~\ref{fig:1} for the choice of parameters $\omega_z = 1.5\omega_c$ and $U=0$. The effect of $U$ is to increase ($U<0$) or decrease ($U>0$) the radius of the region  of the normal phase in the $(g_1,g_2)$ plane.

\begin{figure*}[t]
        \includegraphics[width = \textwidth]{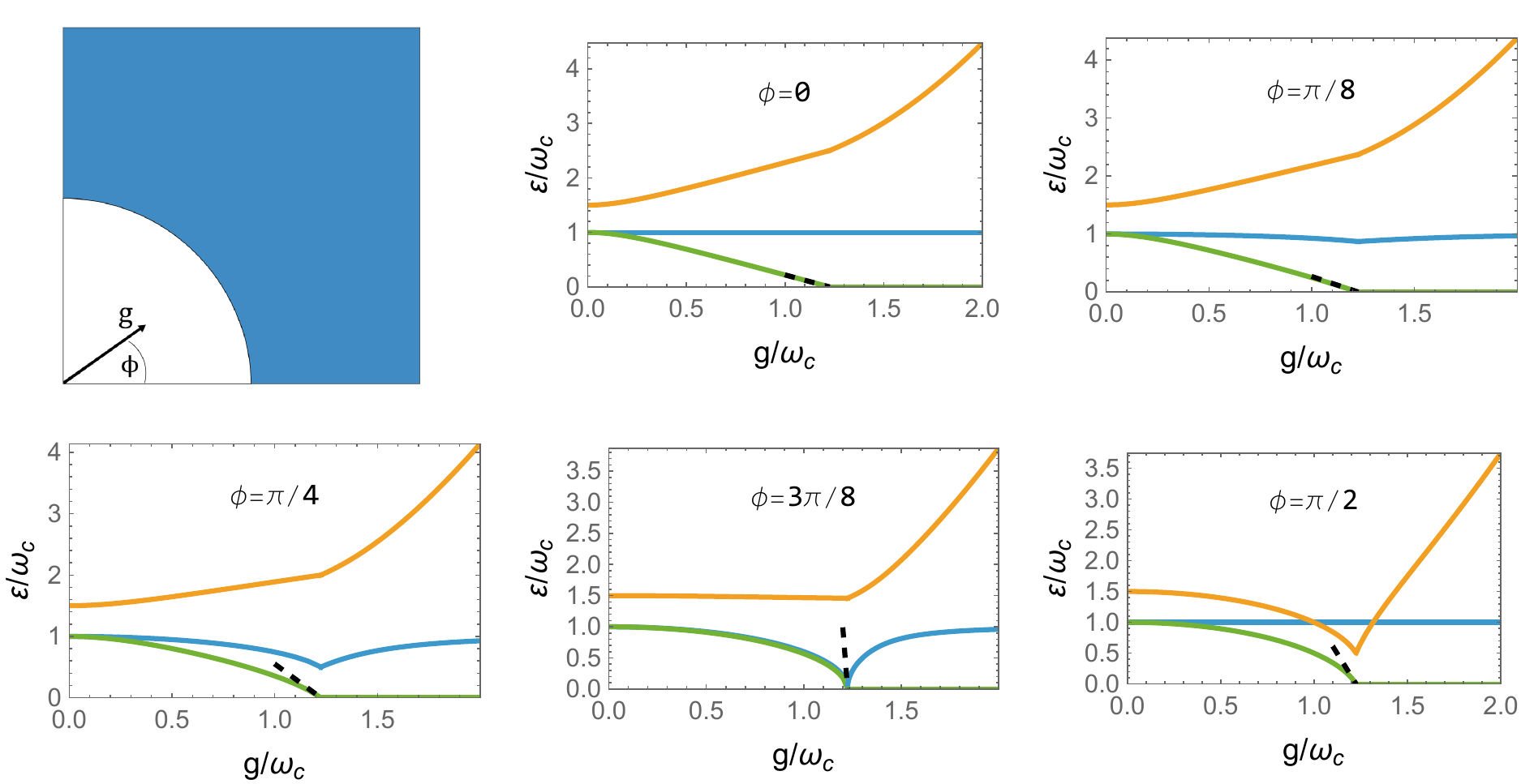}
    \caption{Energy spectrum of Gaussian fluctuations above the mean-field ground state across different cuts through the phase diagram. The dashed black line corresponds to the expression in Eq.~\eqref{eq:linearslope}. The $\phi = 3\pi/8$ trajectory shows that the two lower branches are almost degenerate, resulting in a large slope in Eq.~\eqref{eq:linearslope}. Parameters are $\omega_z=1.5\omega_c$.}
    \label{fig:cuts}
\end{figure*}
{\emph{Spectrum of fluctuations:}} We now proceed to investigate the spectrum of the Gaussian fluctuations on top of the mean-field solution of the ground state. The Bogoliubov Hamiltonian describing these fluctuations takes the form~\cite{bogoliubovSM}: 
\begin{equation}
H_{B} = \sum_{i=1}^3 \sum_{j=1}^3 \Bigg(A_{ij}\tilde{a}^\dagger_{i}\tilde{a}_{j} + \frac{1}{2}B_{i j}\tilde{a}^\dagger_{i}\tilde{a}^\dagger_{j}+\frac{1}{2}B^{\ast}_{i j}\tilde{a}_{i}\tilde{a}_{j}\Bigg).
\end{equation}
The corresponding eigenmodes $\varepsilon_1$, $\varepsilon_2$, $\varepsilon_3$ can be obtained by solving the equation:
\begin{equation}
\det    \begin{pmatrix}
A-\varepsilon \mathbb{I}  &B \\
B^* & A^*+\varepsilon \mathbb{I}
\end{pmatrix} =0,  
\end{equation}
where $\mathbb{I}$ is the $3\times 3$ identity matrix. In the following analysis, we primarily discuss the $U=0$ regime, where compact analytical expressions can be obtained. Further details on the procedure to compute these eigenmodes and the spectrum of fluctuations in the $U\neq0$ regime are discussed in the Supplemental Material (SM)~\cite{SuppMat}.

We begin by noting that the explicit expressions for the eigenmodes are not readily available in the normal phase. Interestingly however, a simple analytical expression for the two dispersive polaritonic modes in the superradiant phase can be obtained~\cite{SuppMat}:
\begin{equation}
\label{eq:pmmodes}
\begin{split}
2\varepsilon_{\pm}^2 = &\,2\omega_c^2+\frac{(g_1^2+g_2^2)^2}{\omega_c^2}+\frac{2\omega_z\omega_c(g_1^2-g_2^2)}{g_1^2+g_2^2}\\
&\pm \sqrt{\frac{(g_1^2+g_2^2)^4}{\omega_c^4}+\frac{4\omega_z(g_1^4-g_2^4)}{\omega_c}+4\omega_z^2\omega_c^2}.
\end{split}    
\end{equation}
Furthermore, the $U(1)$-symmetry breaking leads to the emergence of a gapless Goldstone mode, $\varepsilon_{\Delta}=0$ in the superradiant phase. 

\begin{figure}[t]
        \includegraphics[width = 0.47 \textwidth]{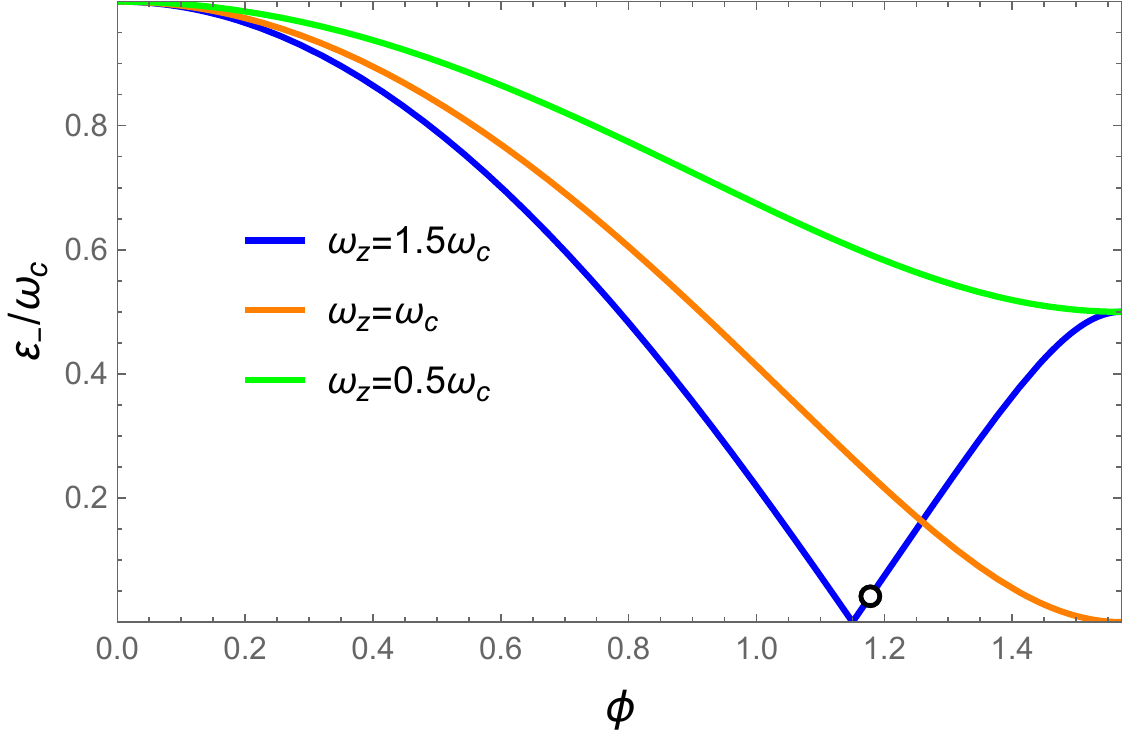}
    \caption{The energy of the lower polariton mode $\varepsilon_{-}$ from Eq.~\eqref{eq:pmmodes} along the critical line $g_{c} = \omega_z\omega_c$ for various values of $\omega_z$. The top curve (green) is always above zero; the middle (orange) curve approaches zero only at $\phi = \pi/2$, corresponding to the anti-Tavis-Cummings limit. The bottom (blue) curve reaches zero when $\phi = 0.5\arccos{(\omega_c/\omega_z)}$. The open marker corresponds to the energy along the cut $\phi = 3\pi/8$ in Fig. \ref{fig:cuts}.}
    \label{fig:lower_branch}
\end{figure}

We present the results for the excitation spectrum both in the normal and the superradiant phase in Fig.~\ref{fig:cuts}  for the choice of the parameters $\omega_z=1.5\omega_c$ and for five different lines in the plane $(g_1,g_2) = (g\cos(\phi),g\sin(\phi))$ that interpolate between the well-studied Tavis-Cummings ($\phi=0$) and anti-Tavis-Cummings ($\phi=\pi/2$) limits. We observe that upon approaching the critical point along four of these lines ($\phi= \{0, \pi/8, \pi/4, \pi/2\}$), the lowest energy branch approaches zero linearly (dashed black curves), suggesting the value of dynamical critical exponent $z\nu=1$. Strikingly, this behavior breaks down along the $\phi = 3\pi/8$ line, for which the slope becomes very steep. This behavior originates from the degeneracy of the two lower energy branches at the critical point, such that the dispersion of the lowest energy branch takes the form: $\varepsilon_{\Delta} \sim|g_{c}-g|^{\frac{1}{2}}$. We demonstrate this explicitly by perturbatively computing the dispersion of the lowest branch in the normal phase, when the energy branches are not degenerate~\cite{SuppMat}:
\begin{equation}
\label{eq:linearslope}
\varepsilon_{\Delta} \approx \frac{2g_{c}}{|\omega_c+\cos(2\phi)\omega_z|}|g_{c}-g|.    
\end{equation}
This perturbative calculation breaks down when $\omega_c = -\cos(2\phi)\omega_z$, and the two lowest branches become degenerate. As demonstrated in Fig.~\ref{fig:lower_branch}, this degeneracy will always occur as long as $\omega_0\geq\omega$, and for the parameters $\omega_z = 1.5\omega_c$ (Fig.~\ref{fig:cuts}), this degeneracy occurs at $\phi = 1.15026$. We verify this conclusion by plotting the energy of the $\varepsilon_{-}$ mode in the superradiant phase given by Eq.~\eqref{eq:pmmodes} in Fig.~\ref{fig:lower_branch}. 

\begin{figure}
\includegraphics[width = 0.47 \textwidth]{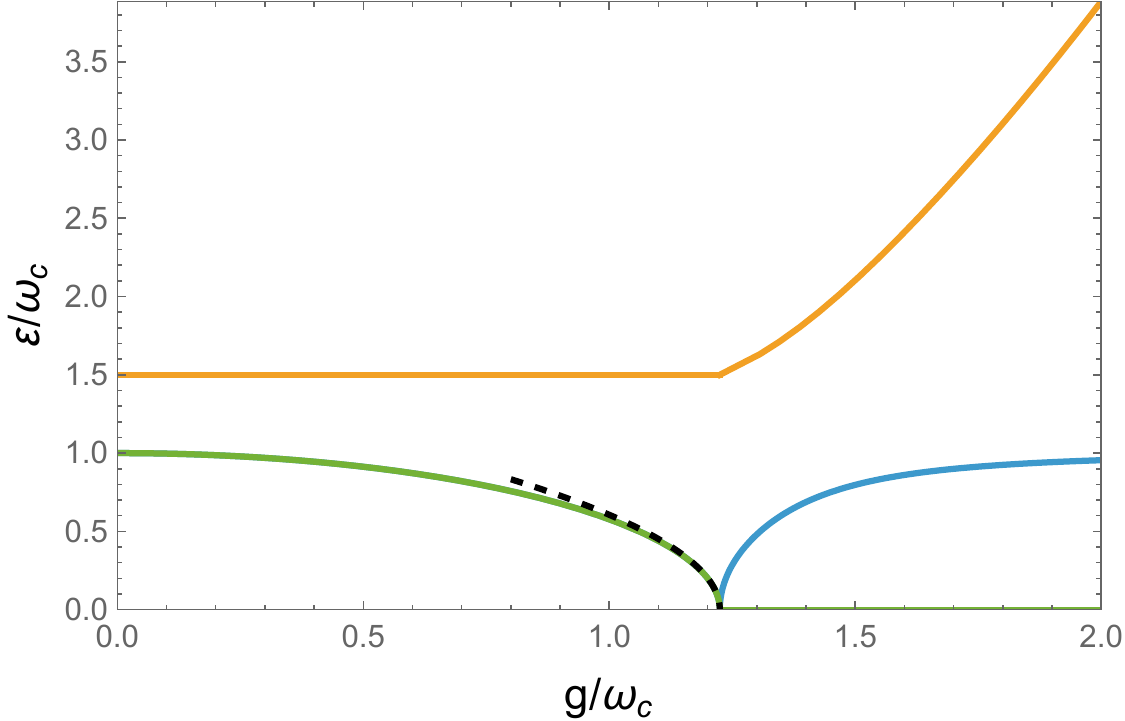}
\caption{The energy branches for the case $\phi = 0.5\arccos{(\omega_c/\omega_z)}$, where two of the branches remain degenerate in the normal phase. The dashed black curve corresponds to Eq.~\eqref{eq:crit}. These results correspond to $\omega_z = 1.5\omega_c$.}
\label{fig:degenerate_modes}
\end{figure}

As detailed in the SM~\cite{SuppMat}, we can extract the gap scaling at the degeneracy point $\cos(2\phi) = -\omega_c/\omega_z$ from the roots of the characteristic equation: $\varepsilon_1 = \omega_z$, $\varepsilon_{2}=\varepsilon_3 = \sqrt{\omega_c^2-g^2\omega_c/\omega_z}$. This implies that at this point the gap closes in the following manner:
\begin{equation}
\label{eq:crit}
\varepsilon_{\Delta}\approx \frac{\sqrt{2}\omega_c}{\sqrt{g_{c}}}|g_{c}-g|^{\frac{1}{2}}.
\end{equation}
As illustrated in Fig.~\ref{fig:degenerate_modes}, our analytical calculations match the exact numerical results quite accurately.

Our analysis suggests that when $\omega_z<\omega_c$, the dynamical critical exponent will be $z\nu=1$, which suggests that the chiral Dicke model lies in the same universality class as the  $SO(2)$ Lipkin-Meshkov-Glick(LMG)  model in this regime~\cite{defenuRMP}. By contrast, for $\omega_z>\omega_c$, there will be a line in the phase diagram along which the dynamical critical exponent is $z\nu=1/2$ akin to the standard DM, which the superradiant phase spontaneously breaks the discrete $\mathbb{Z}_2$ symmetry~\cite{emary&brandespre}. We conclude that different critical exponents emerge in this system depending on the path along which the critical surface is traversed. Thus, the chiral DM provides a route to realize multiversality in atom-photon coupled systems~\cite{multiversality}.

Before concluding our analysis, we briefly discuss the fate of the system when $U\neq 0$. In this regime,  
the Hamiltonian governing the fluctuations in the normal phase takes the same form as the $U=0$ case, with $\omega_c\to \tilde{\omega}_c$. Consequently, our previous analysis of the structure of the modes remains unchanged. Notably, the two degenerate normal phase modes and the consequent multiversality persists in the $\omega_0>\tilde{\omega}$ regime. Analogously, the superradiant phase also hosts a single Goldstone mode and two dispersive modes. Unlike the $U=0$ case however, it is difficult to obtain a simple analytical expression for the dispersive modes $\varepsilon_{\pm}$. Finally, we note that when $g_1$ or $g_2$ is set to zero, the corresponding mode remains dispersive, unlike the $U=0$ regime, where it completely decouples from the system. This feature originates from the renormalization of the cavity mode frequency $\omega_c \to \tilde{\omega}_c+U|\alpha_3|^2$ \cite{SuppMat}.\\

\emph{Conclusions and outlook:} 
In this Letter, we introduced the chiral DM, extending the concept of chiral light-matter coupling as the single-qubit two-mode Rabi model to the many-body regime. By mapping the ground-state phase diagram, we demonstrated a transition from a U(1)-symmetric normal phase to a U(1)-broken superradiant phase. Crucially, this continuous symmetry breaking is an inherent feature of the model, occurring over a broad parameter space---a sharp departure from other generalized DMs where such symmetries are typically fragile and require fine-tuning. We further characterized the tunability of the fluctuation spectrum across both phases. Most significantly, our analysis reveals that the normal-to-superradiant transition exhibits multiversality, where the dynamical critical exponent depends on the specific trajectory in parameter space. Our results suggest that chiral interactions provide a robust mechanism for engineering non-universal critical phenomena in atom-photon systems.


Several promising avenues for future research emerge from this work. The first natural step would be to examine the effect of dissipation on this system~\cite{damanet2019atom,muller2025genuine,mondal2026transient}. Another promising direction would be to characterize the dynamical phases that emerge in this system under periodic~\cite{gong2018discrete,zhu2019dicke,das2024discrete,jager2024dissipative,cosme2023bridging} and quasi-periodic driving~\cite{das2023periodically,anisur2026dissipative}.  Finally, extending this framework to spin chains coupled to two-mode cavities may reveal novel non-equilibrium dynamics and exotic quantum phases~\cite{dickeising2020,KPS2020,KPS2025prb,li2020fast,li2021long,roman2025bound}.

\vspace{5mm}
\begin{acknowledgments}
This work was supported by AFOSR Grant No.~FA9550-23-1-0598 (NY and WVL). SC  thanks Abhishodh Prakash for introducing him to the concept of multiversality.
\end{acknowledgments}

\bibliography{references}

\onecolumngrid
\begin{center}
\newpage
\textbf{
Supplemental Material:\\[4mm]
\large Robust continuous symmetry breaking and multiversality in the chiral Dicke model \\ }

\vspace{4mm}
Nikolay Yegovtsev,$^1$ Sayan Choudhury,$^{2,3}$ W. Vincent Liu$^1$ \\
\vspace{2mm}
{\em \small
$^1$Department of Physics and Astronomy and IQ Initiative, University of Pittsburgh, Pittsburgh, Pennsylvania 15260, USA
$^2$Harish-Chandra Research Institute, a CI of Homi Bhabha National Institute, Chhatnag Road, Jhunsi, Allahabad 211019
$^3$Homi Bhabha National Institute, Training School Complex,
Anushakti Nagar, Mumbai 400 094, India
}
\end{center} 

\setcounter{equation}{0}
\setcounter{figure}{0}
\setcounter{table}{0}
\setcounter{section}{0}
\setcounter{page}{1}
\makeatletter
\renewcommand{\theequation}{S.\arabic{equation}}
\renewcommand{\thefigure}{S\arabic{figure}}
\renewcommand{\thetable}{S\arabic{table}}
\renewcommand{\thesection}{S.\arabic{section}}
\renewcommand{\theHequation}{S.\arabic{equation}}
\renewcommand{\theHfigure}{S\arabic{figure}}
\renewcommand{\theHtable}{S\arabic{table}}
\renewcommand{\theHsection}{S.\arabic{section}}

\section{Analysis of fluctuations}
\subsection{General Hamiltonian}
The general form of the Hamiltonian governing Gaussian fluctuations on top of the mean-field solution is given by:
\begin{equation}
\label{eq:HfiniteU}
\begin{split}
H =&\, \left(\tilde{\omega}_c+U|\alpha_3|^2\right)(\tilde{a}_1^\dagger \tilde{a}_1+\tilde{a}_2^\dagger \tilde{a}_2)+\left[\omega_z+U\frac{(g_1^2+g_2^2)|\alpha_3|^2(N-|\alpha_3|^2)}{N(\tilde{\omega}_c+U|\alpha_3|^2)^2}+\frac{(g_1^2+g_2^2)|\alpha_3|^2(4N-3|\alpha_3|^2)}{2(\tilde{\omega}_c+U|\alpha_3|^2)N(N-|\alpha_3|^2)}\right]\tilde{a}_3^\dagger \tilde{a}_3    \\
&+\frac{g_1}{\sqrt{N}}\left[\frac{(2N-3|\alpha_3|^2)}{2\sqrt{N-|\alpha_3|^2}}-\frac{U|\alpha_3|^2\sqrt{N-|\alpha_3|^2}}{\tilde{\omega}_c+U|\alpha_3|^2}\right](\tilde{a}_1^\dagger \tilde{a}_3+\tilde{a}_3^\dagger \tilde{a}_1)\\
&-\frac{g_1}{\sqrt{N}}\left[\frac{|\alpha_3|^2}{2\sqrt{N-|\alpha_3|^2}}+\frac{U|\alpha_3|^2\sqrt{N-|\alpha_3|^2}}{\tilde{\omega}_c+U|\alpha_3|^2}\right](e^{2i\theta}\tilde{a}_1^\dagger \tilde{a}_3^\dagger + e^{-2i\theta}\tilde{a}_1\tilde{a}_3)\\
&+\frac{g_2}{\sqrt{N}}\left[\frac{(2N-3|\alpha_3|^2)}{2\sqrt{N-|\alpha_3|^2}}-\frac{U|\alpha_3|^2\sqrt{N-|\alpha_3|^2}}{\tilde{\omega}_c+U|\alpha_3|^2}\right](\tilde{a}_2^\dagger \tilde{a}_3^\dagger +\tilde{a}_2\tilde{a}_3)\\
&-\frac{g_2}{\sqrt{N}}\left[\frac{|\alpha_3|^2}{2\sqrt{N-|\alpha_3|^2}}+\frac{U|\alpha_3|^2\sqrt{N-|\alpha_3|^2}}{\tilde{\omega}_c+U|\alpha_3|^2}\right](e^{-2i\theta}\tilde{a}_2^\dagger \tilde{a}_3 + e^{2i\theta}\tilde{a}_3^\dagger \tilde{a}_2)\\
&+\frac{(g_1^2+g_2^2)|\alpha_3|^2(2N-|\alpha_3|^2)}{4(\tilde{\omega}_c+U|\alpha_3|^2)N(N-|\alpha_3|^2)}(e^{2i\theta}\tilde{a}_3^{\dagger 2}+e^{-2i\theta}\tilde{a}_3^2).
\end{split}    
\end{equation}
We obtain the Hamiltonian in the normal and the superradiant state by inserting $|\alpha_3|^2=0$ and $|\alpha_3|^2=\frac{\tilde{\omega}_c}{U}\left(\sqrt{\frac{\tilde{\omega}_c+UN}{\tilde{\omega}_c+\tilde{\mu}UN}}-1\right)$, where $\mu = \omega_z\tilde{\omega}_c/(g_1^2+g_2^2)$ respectively. We note that in the $U\to 0$ regime, we substitute $\tilde{\omega}_c\to \omega_c$, and $|\alpha_3|^2 = \frac{N}{2}\left(1-\frac{\omega_z\omega_c}{g_1^2+g_2^2}\right)$.

\subsection{Diagonalization procedure for $U=0$}
We now outline the procedure to diagonalize the Bogoliubov Hamiltonian when $U=0$. In this case, the most general quadratic Hamiltonian can be written in the form:
\begin{equation}
H_{B} = \sum_{i=1}^3 \sum_{j=1}^3 \Bigg(A_{ij}\tilde{a}^\dagger_{i}\tilde{a}_{j} + \frac{1}{2}B_{i j}\tilde{a}^\dagger_{i}\tilde{a}^\dagger_{j}+\frac{1}{2}B^{\ast}_{i j}\tilde{a}_{i}\tilde{a}_{j}\Bigg).
\end{equation}
We obtain the eigenmodes by solving the equation:
\begin{equation}
\det    \begin{pmatrix}
A-\varepsilon \mathbb{I}  &B \\
B^* & A^*+\varepsilon \mathbb{I}
\end{pmatrix} =0,  
\end{equation}
where $\mathbb{I}$ is the $3\times 3$ Identity matrix. We expand the original Hamiltonian Eq.~\eqref{eq:Hcdm} up to quadratic order in the creation and annihilation operators around the normal state $\alpha_1=\alpha_2=\alpha_3$, and obtain the following expressions of matrices $A_N$ and $B_N$:
\begin{equation}
A_N = \begin{pmatrix}
\omega_c & 0 & g_1\\
0 &\omega_c & 0\\
g_1 & 0 & \omega_z
\end{pmatrix}, \hspace{5mm}
B_N = \begin{pmatrix}
0 & 0 & 0\\
0 & 0 & g_2\\
0 & g_2 & 0
\end{pmatrix}.
\end{equation}
Diagonalization results in the following equation:
\begin{equation}
\label{eq:3branches}
\varepsilon^6 - \left[\omega_z^2+2(\omega_c^2+g_1^2-g_2^2)\right]\varepsilon^4+\left[2\omega_z^2\omega_c^2+(\omega_c^2+g_1^2-g_2^2)^2-2\omega_z\omega_c(g_1^2+g_2^2)\right]\varepsilon^2-\omega_c^2(g_1^2+g_2^2-\omega_z\omega_c)^2=0.    
\end{equation}
There is no simple expression for the roots, but the qualitative behavior of the excitation spectrum in the regime of high imbalance can be inferred from the limiting cases $g_1=0$ or $g_2=0$. Indeed in the case $g_1=0$, mode $a_1$ decouples, and we can divide the equation by $\varepsilon^2-\omega_c^2$ to get:
\begin{equation}
\varepsilon^4-(\omega_z^2+\omega_c^2-2g_2^2)\varepsilon^2+(g_2^2-\omega_z\omega_c)^2=0.
\end{equation}
This yields the following solutions:
\begin{equation}
\varepsilon_{\pm}(g_2) = \sqrt{\frac{\omega_z^2+\omega_c^2-2g_2^2\pm\sqrt{(\omega_z^2-\omega_c^2)^2-4g_2^2(\omega_z-\omega_c)^2}}{2}}    
\end{equation}
When we approach the critical coupling $g_c$, we can expand the $\varepsilon_{-}$ around $g_{c} = \sqrt{\omega_z\omega_c}$ to get:
\begin{equation}
\varepsilon_{-}\approx\frac{2\sqrt{\omega_z\omega_c}}{|\omega_z-\omega_c|}|g_{c}-g_{2}|.  
\end{equation}
We note that both modes become degenerate, when $\omega_z=\omega_c$:
\begin{equation}
\varepsilon_{\pm}(g_2) = \sqrt{\omega_c^2-g_2^2},    
\end{equation}
and near $g_{c} = \omega_c$ the energy of the lower mode scales as:
\begin{equation}
\varepsilon_{-}(g_2) \approx \sqrt{2g_{c}}|g_{c}-g_2|^\frac{1}{2}.    
\end{equation}
Similarly, when $g_2=0$, we obtain:
\begin{equation}
\varepsilon^4-(\omega_z^2+\omega_c^2+2g_1^2)+(g_1^2-\omega_z\omega_c)^2,  
\end{equation}
leading to the following solutions:
\begin{equation}
\varepsilon_{\pm}(g_1) = \sqrt{\frac{\omega_z^2+\omega_c^2+2g_1^2\pm\sqrt{(\omega_z^2-\omega_c^2)^2+4g_1^2(\omega_z+\omega_c)^2}}{2}}.    
\end{equation}
When we approach the critical coupling $g_c$, we can expand the $\varepsilon_{-}$ around $g_{c} = \sqrt{\omega_z\omega_c}$ to obtain:
\begin{equation}
\varepsilon_{-} = \frac{2\sqrt{\omega_z\omega_c}}{(\omega_z+\omega_c)}|g_{c}-g_1|.    
\end{equation}
We can extract how the gap closes as we approach the critical point $g_{c}=\sqrt{\omega_z\omega_c}$ from the normal state by solving Eq.~\eqref{eq:3branches} perturbatively by first introducing polar coordinates $(g,\phi)$ and expressing $g_1=g\cos(\phi)$, $g_2=g\sin(\phi)$ by in powers of $g_{c}-g$:
\begin{equation}
\varepsilon_{\Delta} = \frac{2g_{c}}{|\omega_c+\cos(2\phi)\omega_z|}|g_{c}-g|.
\end{equation}
In the limit where $\phi=0$ or $\phi=\pi/2$, we reproduce the above results obtained from the exact spectrum.

Now, let us consider the degeneracy point $\cos(2\phi) = -\omega_c/\omega_z$. We claim that in this case, two of the modes become degenerate, while the remaining mode has energy $\omega_z$. Indeed, assuming that we have two degenerate roots of the polynomial in Eq.~\eqref{eq:3branches}, we write it as $(\varepsilon^2-\varepsilon_1^2)^2(\varepsilon^2-\varepsilon_2^2)=0$, and compare the coefficients upon expansion:
\begin{equation}
\begin{split}
&\varepsilon_2^2+2\varepsilon_1 ^2 = \omega_z^2+2\left(\omega_c^2-\frac{\omega_c}{\omega_z}g^2\right),\\
&\varepsilon_1^2(\varepsilon_1^2+2\varepsilon_2^2) = 2\omega_z^2\omega_c^2+\left(\omega_c^2-\frac{\omega_c}{\omega_z}g^2\right)^2-2\omega_z\omega_c g^2,\\
&\varepsilon_1^4\varepsilon_2^2 = \omega_c^2(g^2-\omega_z\omega_c)^2.
\end{split}    
\end{equation}
Now it is easy to verify that $\omega_1 = \sqrt{\omega_c^2-g^2\omega_c/\omega_z}$ and $\varepsilon_2 = \omega_z$ solve the above system of equations.

\begin{figure*}[t]
        \includegraphics[width = 1\textwidth,keepaspectratio]{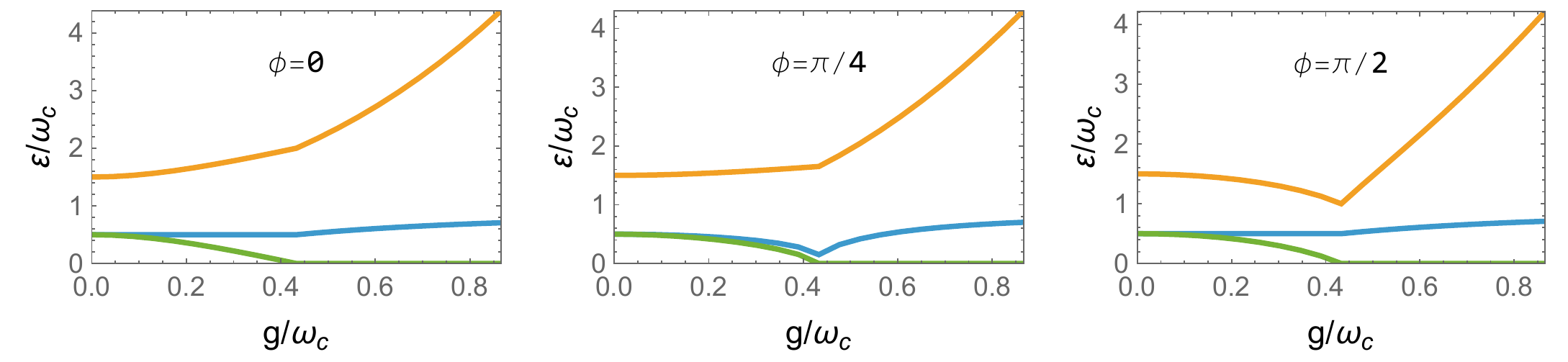}
    \caption{
    The energy branches across different cuts through the phase diagram for the choice of parameters $\omega_z = 1.5\omega_c$ and $UN=1$. This analysis is similar to Fig.~2 in the main text.}
    \label{fig:finiteU}
\end{figure*}

When we insert the expression for the superradiant state $\alpha_3 = |\alpha_3|e^{i\theta} = \sqrt{N/2}\sqrt{(1-\mu)}e^{i\theta}$ with $\mu = \omega_z\omega_c/(g_1^2+g_2^2)$, we obtain the following expressions for matrices $A_S$ and $B_S$:
\begin{equation}
A_S = \begin{pmatrix}
\omega_c & 0 & g_1 (1+3\mu)/(2\sqrt{2}\sqrt{1+\mu})\\
0 & \omega_c & -\bigg(g_2(1-\mu)/(2\sqrt{2}(1+\mu))\bigg)e^{-2i\theta}\\
g_1(1+3\mu)/(2\sqrt{2}\sqrt{1+\mu}) & -\bigg(g_2(1-\mu)/(2\sqrt{2}\sqrt{1+\mu})\bigg)e^{2i\theta} & \omega_z\left(4+(1+\mu)^2\right)/(4\mu(1+\mu))
\end{pmatrix}, \hspace{5mm}
\end{equation}
\begin{equation}
B_S = \begin{pmatrix}
0 & 0 & -\Bigg(g_1(1-\mu)/(2\sqrt{2}\sqrt{1+\mu})\Bigg)e^{2i\theta}\\
0 & 0 & g_2(1+3\mu)/(2\sqrt{2}\sqrt{1+\mu})\\
-\Bigg(g_1(1-\mu)/(2\sqrt{2}\sqrt{1+\mu})\Bigg)e^{2i\theta} & g_2(1+3\mu)/(2\sqrt{2}\sqrt{1+\mu}) & \omega_z(1-\mu)(3+\mu)/(4\mu(1+\mu))
\end{pmatrix}.
\end{equation}
Diagonalization results in the following equation:
\begin{equation}
\varepsilon^6-\left[2\omega_c^2+\frac{(g_1^2+g_2^2)^2}{\omega_c^2}+\frac{2\omega_z\omega_c(g_1^2-g_2^2)}{g_1^2+g_2^2}\right]\varepsilon^4+\frac{(g_1^2+g_2^2)^2(\omega_c^4+(g_1^2+g_2^2)^2)+2\omega_z\omega_c^3(g_1^4-g_2^4)-4\omega_z^2\omega_c^2g_1^2g_2^2}{(g_1^2+g_2^2)^2}\varepsilon^2 =0     
\end{equation}
As anticipated, there is a zero root corresponding to the Goldstone mode. The remaining two modes can be computed analytically:
\begin{equation}
2\varepsilon_{\pm}^2(g_1,g_2) = 2\omega_c^2+\frac{(g_1^2+g_2^2)^2}{\omega_c^2}+\frac{2\omega_z\omega_c(g_1^2-g_2^2)}{g_1^2+g_2^2}\pm\sqrt{\frac{(g_1^2+g_2^2)^4}{\omega_c^4}+\frac{4\omega_z(g_1^4-g_2^4)}{\omega_c}+4\omega_z^2\omega_c^2}.    
\end{equation}
We conclude this discussion by analyzing the finite $U$ regime of Eq.~\eqref{eq:HfiniteU}. To this end, we obtain the phase diagram for $\omega_z=1.5\omega_c$, $UN = 1$ and three choices $\phi = 0$, $\phi = \pi/4$, and $\phi=\pi/2$ numerically. Our results are shown in Fig.~\ref{fig:finiteU}. We observe that when $g_2=0$ ($\phi=0$) or $g_1=0$ ($\phi=\pi/2$), one of the modes remains flat in the normal phase just like in the case of $U=0$, but this mode evolves as $\varepsilon = \tilde{\omega}_c+U|\alpha_3|^2$ in the superradiant phase.

\end{document}